\documentclass[journal]{IEEEtran}
\usepackage[pdftex]{graphicx}
\usepackage{epstopdf}
\usepackage{amsmath, amssymb}
\usepackage{fancyhdr}

\usepackage{subfigure}

\usepackage{adjustbox,lipsum}

\usepackage{color}

\begin{document}

\title{
Applying a formula for generator redispatch to damp interarea oscillations using synchrophasors}

\author{  Sarai Mendoza--Armenta,~\IEEEmembership{Member,~IEEE} \hspace{1cm}Ian Dobson,~\IEEEmembership{Fellow,~IEEE}
\thanks{
The authors are with ECpE dept.,
Iowa State University,
Ames IA USA; dobson@iastate.edu.
We gratefully acknowledge support in part from NSF grant CPS-1135825 and Arend J. and Velma V. Sandbulte professorship funds.
}}
\fancyhead[c]{\textnormal{\small to appear in IEEE Transactions on Power Systems, accepted September 2015}}
\renewcommand{\headrulewidth}{ 0.0pt}
 \fancyfoot[L]{~\\[-20pt]Preprint \copyright 2015 IEEE. \small Personal use of this material is permitted. Permission from IEEE must be obtained for all other uses, in any current or future media, including reprinting/republishing this material for advertising or promotional purposes, creating new collective works, for resale or redistribution to servers or lists, or reuse of any copyrighted component of this work in other works.}
  \fancyfoot[C]{~ }

\maketitle
\thispagestyle{fancy}

\begin{abstract}
\looseness=-1
If an interarea oscillatory mode has insufficient damping, generator redispatch  can 
be used to improve its damping.
We explain and apply a new analytic formula for the modal sensitivity 
to rank the best pairs of generators to redispatch.
The formula requires some dynamic power system data and we show how to 
obtain that data from synchrophasor measurements.
The application of the formula to damp interarea modes is explained and illustrated with
interarea modes of the New England 10-generator power system.
\end{abstract}


\section{Introduction}
\looseness=-1
Power transmission systems have multiple electromechanical oscillatory modes in which 
power system areas 
can swing against each other. In large grids, these interarea oscillations typically have low frequency in the 
range 0.1 to 1.0 Hz, and can appear for large or unusual power transfers.
Poorly damped or negatively damped oscillations can become 
more frequent as power systems experience  greater variability of loading conditions and 
can lead to equipment damage, malfunction or blackouts.
Practical rules for power system security often require sufficient damping of oscillatory 
modes  \cite{Cigre,Rogersbook}, such as damping ratio of at least 5\%, and power transfers on tie lines are sometimes limited
by oscillations \cite{Cigre,IEEE90,IEEE94,ChungPS04}.

There are several approaches to maintaining sufficient damping of oscillatory modes, including 
limiting power transfers \cite{mangoReport}, installing closed loop controls \cite{Cigre}, and the approach of this paper, which is to 
take operator actions such as redispatching generation \cite{mangoReport,DobsonPSERC99,FischerIREP}. 
It is now feasible to monitor modal damping and frequency online from synchrophasors (also known as PMUs) \cite{PierrePS97,WiesPS03}.
Suppose that a mode with insufficient damping ratio is detected.
Then what actions should be taken to restore the mode damping?

This paper calculates the
best generator pairs to redispatch to maintain the mode damping 
by combining synchrophasor and state estimator measurements with a new analytic formula for the sensitivity of the mode eigenvalue
with respect to generator redispatch.
This formula, previously thought to be unattainable, is derived with 
a combination of new and old methods in  \cite{MendozaIREP13}.
The length of the derivation (more than 8 pages) precludes its presentation here.
In this paper we state, explain, and demonstrate the application of the new formula and
show how the terms of the formula could be obtained from power system measurements.
In particular, we propose using synchrophasors to measure the terms of the formula that depend on dynamics and using 
the state estimator to measure the terms of the formula that depend on statics.
In applying a first order sensitivity formula, we assume that the power system ambient or transient 
behavior is dominated by the linearized dynamics associated with an asymptotically stable operating equilibrium.

\looseness=-1
Changes in generator dispatch change the oscillation damping by exploiting  nonlinearity of the power system: changing 
the dispatch changes the operating equilibrium and hence the linearization of the power system about that equilibrium that 
determines the oscillatory modes and their damping. This open loop approach that applies an operator action after too little 
damping is detected can be contrasted with an approach that designs closed loop controls to damp the oscillations preventively. 
The closed loop control design chooses control gains that appear explicitly in the  power system Jacobian, whereas generator 
redispatch changes the Jacobian indirectly by changing the operating point at which the Jacobian is evaluated.

We now review previous works using generator redispatch to damp oscillations;
these have considered heuristics, brute force computations, and formulas that are difficult to implement from measurements.
These approaches have  established that generator redispatch can  damp oscillatory modes.

Fischer and Erlich pioneered heuristics  for the redispatch in terms of the mode shapes for some simple grid structures and for 
the European grid \cite{FischerIREP,FischerPPT}. Their heuristics seem promising for insights, elaborations and validation, 
especially since there has also been progress in determining the mode shape from measurements 
\cite{TrudnowskiPS08,ChaudhuriPS11,DosiekPS13,BarociaPalPS14}. 

There are also previous approaches that require a dynamic grid model. The effective generator redispatches can be determined 
by repetitive computation of eigenvalues  of a dynamic power grid model to give numerical sensitivities 
\cite{ChungPS04,mangoReport,mangoPESGM,DiaoPSCE11}.
Also, there are exact computations of the sensitivity of the damping from a dynamic power grid model 
\cite{DobsonPSERC99,DobsonCDC92,NamPS00,WangJZU08} that are based on the eigenvalue sensitivity formula 
\begin{align}
\label{ReportFormula}
\frac{\partial \lambda}{\partial p} = \frac{wJ_pv}{wv},
\end{align}
\looseness=-1
where $w$ and $v$ are left and right eigenvectors associated with the eigenvalue $\lambda$ and $J_p$ is the derivative 
of the Jacobian with respect to the amount of generator redispatch $p$. The calculation of $J_p$ involves the Hessian 
and the sensitivity of the operating point to $p$.
However, requiring a large scale power system dynamic model poses  difficulties. 
It is challenging to obtain validated models of generator dynamics over a wide area and particularly difficult to 
determine dynamic load models  that would be applicable online when poor modal damping  arises.

We think that a good way to solve the difficulties with online large scale power system dynamic models
is to combine models with synchrophasor measurements to get actionable information about mode damping.
In particular, the dynamic information about the power system can be estimated from synchrophasors.
However, formula  (\ref{ReportFormula}) is not suitable for this purpose since 
it is not feasible  to estimate from measurements the left eigenvector $w$
 in (\ref{ReportFormula}) (or derivatives of eigenvectors in other versions of (\ref{ReportFormula})).
This was a primary motivation for developing our new formula in \cite{MendozaIREP13}.
In particular,  the formula shows that the first order effect of a generator redispatch 
largely depends on the mode shape (the right eigenvector) and  power flow quantities that can be 
measured online. The assumed equivalent generator dynamics only appears as a factor common to all redispatches.
Given a lightly damped interarea mode of a system, the method can rank all the possible generators pairs. 
The rank is based on the size of the change of the damping ratio of the interarea mode. 

The main objective of ranking the generator pairs is to provide advice to the operator of 
 several effective generator redispatches  to damp the oscillations from which a corrective action can be selected.
 There are many economic goals and operational constraints governing the final selection of an appropriate dispatch  
 by the operator,  and the integration of this decision with optimal power flow and 
 the power markets is left to future work.

In this paper, the role in the method of measurements and the formula are first illustrated in a simple 3-bus system. 
Then the general formula derived in \cite{MendozaIREP13} is presented. 
How static and dynamic power quantities are obtained from measurements is discussed.
The complex denominator of the formula contains the effect of the assumed equivalent generator dynamics,
and a technique to obtain from measurements the phase of this complex denominator
is presented. 
The paper then explains how to calculate the best generator pairs to damp a given  mode, and 
illustrates and verifies the calculation with interarea modes of the New England 10-generator system.

\section{\label{Eigenvalue Sensitivity} Formula for eigenvalue sensitivity with respect to generator redispatch}
\subsection{Special case of a 3-bus system}
\begin{figure}[!h]
\includegraphics[width=\columnwidth]{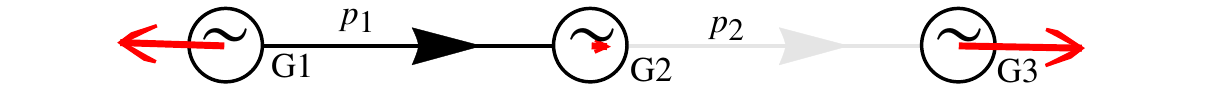}
\caption{\small \label{3BusSystem2}The gray
lines  joining the buses show the magnitude of the  power flow  with the  grayscale 
and the direction of the  power flow 
with the arrows. 
The red arrows at each bus show the oscillation mode shape; 
that is, the magnitude and direction of the 
complex entries of the right eigenvector $x$ associated with
the interarea mode.}
\end{figure}

We first present the eigenvalue sensitivity formula for a special case in which the formula simplifies.
Consider the interarea mode of the simple 3-bus, 3-generator system shown in Fig.~\ref{3BusSystem2}. 
The  generator dynamics is
 described by the swing equation and the transmission lines are lossless.  In this special case, the bus voltage magnitudes are
assumed constant.
At the base case, the system is at an stable
operating point and 
Fig.~\ref{3BusSystem2} shows the line power flows and the interarea oscillating mode pattern.
The mode pattern shows that generator G1 is swinging against G3, and that G2 is not participating in the oscillation.

The bus  voltage phasor angles are $\delta_1,\delta_2,\delta_3$.
Let
$\theta_1=\delta_1-\delta_2$ and $\theta_2=\delta_2-\delta_3$ be the voltage angle differences across line 1 
and line 2, and let $p_1$ and $p_2$ be the real power flows through line 1 and line 2.
The interarea mode has complex eigenvalue $\lambda$ and complex right eigenvector $x$.
Generator redispatch causes changes in the angles across the lines $d\theta_1$ and $d\theta_2$,
and these in turn  cause changes $d\lambda$ in the eigenvalue.
According to \cite{MendozaIREP13}, the formula for the first-order change in  $\lambda$  with respect 
to generator redispatch in a 3-bus system with constant voltage magnitudes reduces to 
\begin{align}
\label{FormulaVConstant}
d\lambda =
\frac{(x'_{\theta_1})^2p_{_{1}}d\theta_1 + (x'_{\theta_2})^2p_{_{2}}d\theta_2}{\alpha}.
\end{align}
All quantities are in per unit unless otherwise indicated. The complex denominator  $\alpha=0.3080\angle 87.12^{\circ}$ 
seconds depends on the inertia and damping coefficients of the generators, the eigenvalue $\lambda$ and its right eigenvector $x$. 
The right eigenvector written with components corresponding to the bus angles is $x=(x_{\delta_1},
x_{\delta_2},x_{\delta_3})^t$.\footnote{Since $x$ is the right eigenvector of a quadratic formulation of the eigenvalue problem \cite{MendozaIREP13},
it contains the angle components, but not the frequency components, of the conventional right eigenvector. That is, the conventional right eigenvector  is 
$(x_{\delta_1},
x_{\delta_2},x_{\delta_3},x_{\omega_1},x_{\omega_2},x_{\omega_3})^t=
(x_{\delta_1},
x_{\delta_2},x_{\delta_3},\lambda x_{\delta_1},
 \lambda x_{\delta_2},\lambda x_{\delta_3})^t=(x,\lambda x)^t$.}
The right eigenvector may also be expressed in terms of the angles across the line as $x'=(x'_{\theta_1},
x'_{\theta_2})^t$. That is, $x'_{\theta_1}=x_{\delta_1}-x_{\delta_2}$ is  the change in the eigenvector across line 1, 
and  $x'_{\theta_2}=x_{\delta_2}-x_{\delta_3}$
is the change of the eigenvector across line 2. It is these changes $x'_{\theta_1}$ and $x'_{\theta_2}$ in the right eigenvector 
across the lines that appear in (\ref{FormulaVConstant}).

\looseness=-1
Formula (\ref{FormulaVConstant}) is in terms of static load flow  quantities $\theta$ and $p$ that are available from state
estimation, and dynamic quantities $\lambda$ and $x$ that could be available from 
synchrophasor measurements. Table \ref{Static and dynamic quantities 3-bus} shows the
values at the base case. Formula (\ref{FormulaVConstant}) indicates which lines have suitable power flow and 
eigenvector components to affect oscillation damping. In particular, it is effective for
the redispatch to change the angles across the lines that have both changes in the 
mode shape across the line and sufficient real power flow in the right direction.

\begin{table}[h]
\caption{\label{Static and dynamic quantities 3-bus}{Quantities from measurements} }
\centering \begin{tabular}{ccc|c}
&  \multicolumn{2} {c|} {Static quantities from}   & Dynamic quantities\\
& \multicolumn{2} {c|}{state estimator} & from synchrophasors\\
Line No. &$\theta$ &  $p$ [pu] & $x'_{\theta}$  \\
\hline
1 & $5.164^{\circ}$ & 0.2 & -0.2958 + j0.009838\\
2 & $2.579^{\circ}$ & 0.1 & -0.2668 + j0.004026\\
\hline
\end{tabular}
\end{table}
Let us define the complex coefficients of $d\theta_1$ and $d\theta_2$ in (\ref{FormulaVConstant}) as 
$C_{\theta_{1}}$ and $C_{\theta_{2}}$ respectively. Then
\begin{align}
\label{Cthetas}
d\lambda&=C_{\theta_{1}}d\theta_1 + C_{\theta_{2}}d\theta_2, \quad\mbox{where}\\
C_{\theta_{1}}=   -0.000928&  - \mbox{j}0.0569 \mbox{,~~} C_{\theta_{2}}=   0.000462  - \mbox{j}0.0231. \notag
\end{align}
Coefficient $C_{\theta_{1}}$ has the largest real and imaginary components. It follows that $d\lambda$ is more
sensitive to redispatches done through line 1 than redispatches done through line 2.
The damping ratio at the base case is given by
\begin{align}
\label{DampingRatioDefinition}
\zeta=-\frac{\mbox{Re}\{\lambda\}}{|\lambda|}=-\frac{\sigma}{|\sigma+\mbox{j}\omega|}=-\frac{\sigma}{\sqrt{\sigma^2+\omega^2}}.
\end{align}
Table \ref{dZeta3BusSysten} shows the results for the three possible generators pairs in the system for a small redispatch of 0.01 pu.
As expected from (\ref{Cthetas}), the damping ratio has the largest changes when redispatch
is done through line 1. This is the case of generator pairs G1+,G2- and G1+,G3-; their changes in $d\zeta$ are close. The damping ratio has 
almost no change when redispatch is done only through line 2 with the pair \mbox{G3+,G2-}.
Gi+,Gj- indicates the redispatch that increases power at generator Gi and decreases power at generator Gj.
\begin{table}[h]%
\caption{\label{dZeta3BusSysten}Generator pairs ranked by change in $\zeta(\%)$; redispatch = 0.01 }
\centering \begin{tabular}{cc}
Generator Pair& $d\zeta(\%)$   \\  
\hline
1  G1+,G2- & 0.000393 \\
2  G1+,G3- &  0.000386 \\
3  G3+,G2- & 0.000007 \\
\hline
\end{tabular}
\end{table}
\subsection{General case of formula for the change in the eigenvalue}
Consider a connected power grid that has  $m$ generators, $n+m$ buses and $\ell$ lines. 
Buses $n+1,\dots,n+m$ are the internal buses of the generators.
Assume AC power flow and lossless transmission lines.
Every generator is modeled with an equivalent second order
equation (swing equation) and constant internal voltage magnitude.
Loads are constant power, but frequency dependence of the real power and voltage dependence of the reactive power can be accommodated \cite{MendozaIREP13}.
Then the power system dynamics is described by a set of
differential-algebraic equations with variables $(\delta,V)$.
The dynamic variables are
$\delta_{n+1},\ldots,\delta_{n+m}$, and the algebraic variables are
$\delta_1,\ldots,\delta_n,V_1,\ldots,V_n$.
 In this paper
 we use the quadratic formulation of the
  eigenvalue problem \cite{MendozaIREP13,EliassonHillPS94,MalladaCDC11} with right eigenvector $x$ and eigenvalue $\lambda$:
 \begin{align}
 \label{quadratic}
 (M\lambda^2+D\lambda+L)x=0,
\end{align}
where $L$ is part of the system Jacobian described in \cite{MendozaIREP13} and  $M$ and $D$ are diagonal matrices containing 
the generator equivalent inertias and dampings; $M=\mbox{diag}\{2h_{1}/\omega_0,2h_{2},/\omega_0,\ldots,2h_{m}/\omega_0,0,\ldots,0\}$
and $D=\mbox{diag}\{d_{1},d_{2},\ldots,d_{m},0,\ldots,0\}$. 
The components of the right eigenvector or mode shape are 
$x=(x_{_{\delta_{_{n+1}}}},\ldots,x_{_{\delta_{_{n+m}}}},x_{_{\delta_{_{1}}}},\ldots,x_{_{\delta_{_{n}}}},
x_{_{V_{_{1}}}},\ldots,x_{_{V_{_{n}}}})^t$.

In other papers, the eigenstructure of differential-algebraic models with extended Jacobian $J$ is  often  analyzed using the 
generalized eigenvalue problem $Jv=\lambda E v$ \cite{SmedPS93}. 
The difference between $x$ and $v$ is that $x$ does not include the components of $v$
related to generator angular speeds.

According to \cite{MendozaIREP13}, the new formula for the sensitivity of a nonresonant (algebraic multiplicity one) 
eigenvalue $\lambda$ 
is 
\begin{align}
\label{dlambda}
d\lambda=\frac{1}{\alpha}\left(\sum_{i=1}^{\ell}\tilde{C}_{\theta_{k}}d\theta_k +\sum_{i=1}^{n}\tilde{C}_{V_i}dV_{i}\right).
\end{align}
The denominator of (\ref{dlambda}), which is the same for all redispatches of a given mode, is  
\begin{align}
\label{alpha}
\alpha=2\lambda x^TMx +x^TDx,
\end{align}
$d\theta_k$ is the change in angle across the line $k$ and
$dV_{i}$ is the change in load $i$ voltage magnitude due to the redispatch.
\begin{align}
\tilde{C}_{\theta_k}&=
[ (x'_{\theta_k})^2-(x'_{\nu_k})^2]p_{_k} + 2x'_{\theta_k}x'_{\nu_k}q_{_k}, 
\label{Cthetak}
\\
\tilde{C}_{V_{i}}&=
\sum_{k=1}^{\ell}|A_{ik}|(-C_{q_{_{k}}}q_{_k}- C_{p_{_{k}}}p_{_k})V_i^{-1}-C_{Q_i}Q_iV_i^{-1},
\label{CVi}\\
\nonumber
C_{q_{_{k}}} & = x'_{\nu_k}\left(x'_{\nu_k}-2\frac{x_{_{V_{i}}}}{V_i} \right) - (x'_{\theta_k})^2,\\
\nonumber
C_{p_{_{k}}} & = 2x'_{\theta_k}\left(x'_{\nu_k} - \frac{x_{_{V_{i}}}}{V_i}\right),
\qquad C_{Q_i}  = -2\left(\frac{x_{_{V_{i}}}}{V_i}\right)^2,\\
p_k & =b_{k}V_iV_j\sin{\theta_k},\qquad\qquad q_k=-b_{k}V_iV_j\cos{\theta_k},\notag
\end{align}
where
\begin{align}
\nonumber
\theta_k &=\left\{\begin{array}{ll}
\delta_i - \delta_j & \text{if bus $i$ is sending end of line $k$}\\
\nonumber
\delta_j - \delta_i &\text{if bus $i$ is receiving end of line $k$}
\nonumber
\end{array}\right. \notag\\
x'_{\theta_k}&=\left\{\begin{array}{ll}
x_{_{\delta_i}}-x_{_{\delta_j} }& \text{if bus $i$ is sending end of line $k$}\\
x_{_{\delta_j}}-x_{_{\delta_i} }&\text{if bus $i$ is receiving end of line $k$}
\end{array}\right. \notag\\
x'_{\nu_k} &=\left\{\begin{array}{ll}
\!\displaystyle\frac{x_{\scriptscriptstyle {V_i}}}{V_i} + \frac{x_{_{V_j}}}{V_j} 
&\text{\hspace{-2mm}if line $k$ joins load bus $i$ to load bus $j$}\\[4mm]
\!\displaystyle\frac{x_{\scriptscriptstyle {V_i}}}{V_i}&
\text{\hspace{-11mm}
if line $k$ joins load bus $i$ to generator bus $j$}
\end{array}\right.\notag
\end{align}
$p_k$ is the real power flow in line $k$, $q_k$ is part of the reactive power flow 
in line $k$, and $|A_{ik}|$ is the absolute value of the $ik$-component of the bus-line
incidence matrix. $Q_i$ is the reactive power demanded by load bus $i$. 

Formula (\ref{dlambda}) expresses the first-order change in the eigenvalue $d\lambda$  in terms of the changes $d\theta$ in angles across the lines 
 and changes $dV$ in load voltage magnitudes  caused by the generator redispatch.
$d\lambda$ depends linearly on the active power flow, part of the reactive power flow
through every line of the network, and the reactive power demands at the loads, all evaluated at the base case.
\subsection{Generator modeling}
\label{genmodeling}
\looseness=-1
The overall dynamics of each generator is  described by an equivalent swing equation model 
with constant internal voltage magnitude. 
For generator bus $i$, an ``equivalent
swing equation" is the standard swing equation model with  inertia and damping coefficients $h_i$ and $d_i$  that 
produce  the second order model that best approximates the entire dynamics of the generator $i$ and its controls
(this would be described as a second order dominant pole approximation in automatic controls).
In our approach, there is no need to model the parameters $h_i$ and $d_i$ for each generator;
it is only necessary to assume the existence of a second order model that can approximate well enough the contribution of the generator to the electromechanical mode.
Indeed, since  (\ref{dlambda}) only includes the combined generator dynamics as a common factor
that is the same for all redispatches, we do not need to know the individual parameters of each equivalent generator model.
Other authors using synchrophasor measurements  to identify aggregated generator dynamics for studying oscillations 
also assume swing equation generator models but for their purposes identify the individual inertia and damping coefficients \cite{ChakraborttyISGT2010,ZhouPNNLNAPS11}.
However, we have to consider the generator modeling independently of previous studies since our generator redispatch application has novel and different modeling requirements as discussed in 
section \ref{discussgenmodels}.

\section{\label{Measurements} Measurements}
Formula (\ref{dlambda}) depends on power system quantities that could be 
observed from measurements from the state estimator and from synchrophasors.
\subsection{Quantities obtained from state estimator: $p,q,Q,d\theta,dV$}
 The state estimator can determine the active power $p$ and part of the reactive power flow $q$ through every line of the 
network, and the load reactive power injections $Q$.
Load flow equations can be used to relate the generator redispatches to changes 
in angles across the lines $d\theta$ and load voltage magnitudes $dV$. 
\subsection{Quantities obtained from synchrophasors: $\lambda,x$}
Formula (\ref{dlambda}) depends on the  dynamic quantities $\lambda$ and $x$
that satisfy the quadratic formulation of the eigenvalue problem (\ref{quadratic}). 
For an electromechanical oscillation present in a system, 
synchrophasors can make online measurements 
\cite{PierrePS97,WiesPS03,VanfrettiIREP10,IEEE12} 
of the damping and frequency of the eigenvalue $\lambda$ associated with the oscillation.
$x$ is easy to obtain from a conventional right eigenvector.
The right eigenvector of $\lambda$ is in principle, and to
some considerable extent in practice, available from ambient
or transient synchrophasor measurements  
\cite{TrudnowskiPS08,ChaudhuriPS11,DosiekPS13}.
It is conceivable that historical observations or offline computations
or general principles about the mode shape could be used to augment 
or interpolate the real-time observations, or that the real-time observations
could be used to verify a predicted mode shape. Thus some combination
of measurements and calculation  could yield the mode shape $x$.

\subsection{\label{Denominator}Method for estimating phase of $ \alpha$ from measurements}
The complex denominator $\alpha=2\lambda x^TMx + x^TDx$ is the only part of 
formula (\ref{dlambda}) that depends on the generator equivalent dynamic parameters in the  inertia and damping matrices $M$ and $D$.
In estimating the change $d\lambda$  in the eigenvalue 
from (\ref{dlambda}), the angle $\angle\alpha$ controls the direction of $d\lambda$ and the size of $\alpha$ controls the size of $d\lambda$. For each mode,  $\alpha$ is the same for all generator redispatches.
Therefore, in order to rank the generator redispatches for a given mode, it is sufficient to estimate $\angle\alpha$ from measurements.

Formula (\ref{dlambda}) can be summarized as
\begin{align}
\label{dlambda Alpha Computation}
d\lambda = \frac{\mbox{Numerator}}{\alpha},\quad \mbox{so}\quad \angle \alpha = \angle \mbox{Numerator} - \angle d\lambda.
\end{align} 
\looseness=-1
To estimate $\angle \alpha$ we propose to take advantage of the small random load variations
around the operating equilibrium.
For such small random load variations, samples of $d\lambda$ could be obtained from  a series of synchrophasor estimates of $\lambda$.
Samples of $d\theta$ and $dV$ can be obtained  from the load flow equations with simulated 
random load variations, so that samples of the numerator of (\ref{dlambda}) can be computed.
It turns out that both the $d\lambda$ samples and numerator samples have preferred or principal directions in the complex plane. 
Analyzing the $d\lambda$  and numerator  samples with Principal Component Analysis gives a principal axis direction for 
$d\lambda$ and a principal axis direction for the numerator, and according to (\ref{dlambda Alpha Computation}), the angle between 
these principal axis directions can be used to find $\angle\alpha$.
In section \ref{Estimating Alpha For New England}, we illustrate and apply this calculation of $\angle\alpha$ to  interarea 
modes of the  New England system. 

\subsection{Measurement processing requirements}

To be able to supply the dynamic quantities in formula (\ref{dlambda}), online synchrophasor monitoring of the oscillatory mode eigenvalue $\lambda$ and mode shape $x$  is needed. 
The modal eigenvalue is used directly in (\ref{dlambda}) and, according to the suggested approach in section \ref{Denominator}, for estimating the phase of $\alpha$. 
This subsection comments briefly about the likely measurement processing limitations. 
We would expect to be able to use standard sampling rates and data concentration similar to that already in use for mode monitoring.

Methods to estimate oscillatory modes and mode shapes from synchrophasor data are deployed and improving, and some recent methods \cite{ChaudhuriPS11,LiuManiPESGM12,NingManiPS13,KhaliliniaManiPD15}, 
have used up to 5 minutes of ambient or transient data for multiple parallel algorithms to converge to consistent results. 
There is also consistency between results from methods for transient response after a disturbance and ambient methods. 
However, it does not matter for our algorithm 
whether the dynamics is estimated from transient or ambient responses.

The mode shape estimation currently samples the mode shape at multiple spatial locations.
We require the mode shape at other locations to be interpolated, perhaps guided by previously 
observed mode shapes for specific modes.
Our approach also requires static quantities to be estimated with standard state estimation.
Given state estimator convergence, the estimated state should be readily available within the several minute time scale already
required for mode estimation. We do not anticipate  significant computational delay in evaluating formula (\ref{dlambda}) and  ranking the generators.

\section{\label{Relating} Relating redispatch to the changes $d\theta$, $dV$, $d\lambda$}
Formula (\ref{dlambda}) expresses  the eigenvalue change $d\lambda$ in terms of  $d\theta$ and $dV$.
It remains to express $d\theta$ and $dV$ in terms of the redispatch $dP$.
Define the coefficient vectors of $d\theta$ and $dV$ in (\ref{dlambda}) as 
\begin{align}
\label{newC}
C_{\theta}=\frac{1}{\alpha}\tilde{C}_{\theta} \qquad \mbox{ and}\qquad C_{V}=\frac{1}{\alpha}\tilde{C}_{V}.
\end{align}
Then  
\begin{align}
\label{dlambaCcoefficients}
d\lambda=C_{\theta}\cdot d\theta + C_{V}\cdot dV
=
(C_{\theta},C_{V})
\begin{pmatrix}
d\theta\\
dV
\end{pmatrix}.
\end{align}
From \cite{MendozaIREP13} we know that the linear relationship between
the changes in angles $d\theta$ and changes in voltages $dV$ with  redispatch 
$dP$ is given by
\begin{align}
\label{dthetadVandPrelation}
\begin{pmatrix}
d\theta\\
dV
\end{pmatrix}
=
\begin{pmatrix}
A^T_{\ell \times (n+m)} & 0_{\ell \times n}\\[4pt]
0_{n\times (n+m)} & I_{n \times n }
\end{pmatrix}
 L^{\dagger}
\begin{pmatrix}
dP\\
0
\end{pmatrix},
\end{align}
where 
$A$ is the network incidence matrix, and $\dagger$
indicates pseudoinverse.
As an alternative to the linearized computation of $( d\theta,
dV)$ from the redispatch $dP$ in (\ref{dthetadVandPrelation}), one can simply recompute the AC load flow 
with the  assumed change in redispatch $dP$.
We can relate the change in an eigenvalue of a mode to a generator redispatch $dP$ 
by substituting (\ref{dthetadVandPrelation}) into (\ref{dlambaCcoefficients}):
\begin{align}
\label{dlambda2}
d\lambda
=C_P\cdot dP=\sum_{i=1}^mC_{P_{i}}dP_{i}.
\end{align}
The complex coefficient $C_{P_{i}}$ gives the contribution of generator $i$ to $d\lambda$ by increasing the real 
power of generator $i$  by  $dP_i$. The redispatch is assumed to satisfy the active power balance constraint 
$\sum_{i}^{m}dP_i =0$. It is convenient to define $\tilde{C}_P=|\alpha|C_P$. Then multiplying both sides of  (\ref{dlambda2}) 
by $|\alpha|$ gives 
\begin{align}
\label{dlambda3}
|\alpha|d\lambda
=\tilde{C}_P\cdot dP=\sum_{i=1}^m \tilde{C}_{P_{i}}dP_{i}.
\end{align}
It can be seen by considering (\ref{newC})--(\ref{dlambda2})
that $\tilde{C}_P$ can be calculated from $\tilde{C}_{\theta}$, $\tilde{C}_{V}$ and $\angle\alpha$.
(Note, for example,  that $|\alpha|C_{\theta}=e^{-\angle\alpha}\tilde{C}_{\theta}$.)
Since $|\alpha| d\lambda$ is $ d\lambda$ multiplied by a real constant, the complex number $|\alpha| d\lambda$ 
has the same  angle as $d\lambda$ and has magnitude proportional to $ d\lambda$.
It follows that we can use $|\alpha| d\lambda$ to rank the generator redispatches.

\section{\label{Ranking generator pairs} Ranking generators pairs by increase in damping ratio  due to redispatch}

For a given oscillatory mode with insufficient damping ratio, the first step towards 
ranking the best generators to redispatch is to make the following measurements and calculations:
\begin{enumerate}
\item The eigenvalue $\lambda$ associated with the interarea oscillation and its 
mode
shape $x$ are estimated from synchrophasor measurements. 
\item $p,q,Q$ are obtained from the state estimator.
\item $\angle \alpha$ is computed from measurements.
\item Coefficients $\tilde{C}_{\theta_k}$ and $\tilde{C}_{V_{i}}$ in (\ref{dlambda}) are computed from  (\ref{Cthetak}) and (\ref{CVi}).
\item $\tilde{C}_P$ is computed from $\tilde{C}_{\theta}$, $\tilde{C}_{V}$,  $\angle\alpha$.
\end{enumerate}
The most straightforward way of implementing generator redispatch 
is with a pair Gi+,Gj- of generators; that is, increasing the real power 
of generator $i$ by some amount $dP_i$ and increasing the real power of generator $j$ by 
the negative  amount $dP_j=-dP_i$.
We can use (\ref{dlambda2}) to find the generator pair Gi+,Gj- that gives 
the most favorable eigenvalue change $d\lambda$ that best increases the damping ratio.
For a redispatch of amount $dP_i$ with generator pair Gi+,Gj-,
the resulting eigenvalue change satisfies 
\begin{align}
\label{dlambda2Pairs}
|\alpha|d\lambda_{ij}= (\tilde{C}_{P_{i}}-\tilde{C}_{P_{j}})dP_i.
\end{align}
Also the opposite redispatch Gj-,Gi+ with the same generator pair will yield 
$|\alpha|d\lambda_{ji}=-|\alpha|d\lambda_{ij}$. 
For every  pair of generators in the 
system we can compute $|\alpha|d\lambda_{ij}$ and $|\alpha|d\lambda_{ji}$ and then
we can compute the corresponding damping ratios $\zeta_{ij}$ and $\zeta_{ji}$  for every generator pair
with  (\ref{DampingRatioDefinition}).
 The damping ratios for all pairs 
$\{\zeta_{12},\zeta_{21},\ldots,\zeta_{m-1,m},\zeta_{m,m-1}\}$ are then ranked 
 from the largest one to the smallest one. The highly ranked pairs
 indicate the generator pairs and redispatch directions
that will increase the damping ratio of the mode the most. 

\section{\label{New England system} Damping modes of the New England system}
This section presents  the results for damping interarea modes of the New England 10-generator
system \cite{PaiBook89} with generator redispatch.
The equivalent parameters of the generators are given in Appendix A
and the rest of the data is provided in \cite{PaiBook89}. All the numerical computation was done with the software Mathematica. 
Table \ref{New England interarea mode eigenvalues} shows the eigenvalues of the 4 interarea modes  at the base case.
Since the system has 10 generators, there are 45 generators pairs.
The method presented in section \ref{Ranking generator pairs} was implemented for the interarea modes
(this computation used a calculated value of $\alpha$ and a recalculated load flow to evaluate $d\theta$ and $dV$ from $dP$).
\begin{table}[h!]
\caption{\label{New England interarea mode eigenvalues}{Interarea mode eigenvalues at the base case} }
\centering
 \begin{tabular}{cccc}
Mode  & Eigenvalue $\lambda$ [1/s] & f [Hz] & Damping Ratio $\zeta$(\%) \\
\hline
 1 & -0.040336 + j3.4135& 0.54327 & 1.18157\\
 2 & -0.018839 + j4.7631 & 0.75807 & 0.39551  \\
 3 & -0.024903 + j5.4994 & 0.87526 & 0.45283  \\
 4 & -0.055799 + j6.0159 & 0.95746 & 0.92748 \\
\hline
\end{tabular}
\end{table}

\subsection{\label{Verifying the formula} Verifying the formula}
Formula (\ref{dlambda}) was verified for the four interarea modes, but
here we present results only for mode 1.
The eigenvalue $\lambda_1$ was computed for a given redispatch using  (\ref{dlambda}) and 
using the exact computation of the Jacobian at the new operating point.
Table \ref{Mode1} shows the exact and approximate $\lambda_{1}$ for
different amounts of redispatch of generator pair \mbox{G5+,G9-}.\footnote{ The base case generations of the generators G1 through G10 
 are $2.5,4.8,6.5,6.32,5.08,6.5,5.6,5.4,8.3,10.0$ per unit respectively. As an example of specifying the redispatch, a 0.03 redispatch of generator 
pair \mbox{G5+,G9-} changes the generation of G5 from 5.08 to 5.11 and the generation of G9 from 8.3 to 8.27.}

Table \ref{Mode1} confirms that (\ref{dlambda}) reproduces
the first order variation of $\lambda_1$ with respect to the redispatch.\newline
\begin{table}[h!]
\caption{\label{Mode1}{Eigenvalue $\lambda_{1}$ for generator redispatch G5+, G9-} }
\centering
\begin{tabular}{ccc}
Redispatch & Exact eigenvalue & Approximate eigenvalue \\
\hline
0.0 & -0.0403355 + j3.4135  & -0.0403355 + j3.4135  \\
0.0005 & -0.0403225 + j3.4131 & -0.0403225 + j3.4131  \\
0.001 & -0.0403095 + j3.4127  & -0.0403095 + j3.4127  \\
0.01 & -0.0400715 + j3.4059  & -0.0400736 + j3.4059  \\
0.02 & -0.0397985 + j3.3981  & -0.0398086 + j3.3983  \\
0.03 & -0.0395161 + j3.3899  & -0.0395403 + j3.3905  \\
 \hline
\end{tabular}
\end{table}
\subsection{\label{Estimating Alpha For New England} Estimating the phase of $ \alpha$ for random load variations}
Table \ref{AlphaPhase} shows the exact  $\angle \alpha$ computed with (\ref{alpha}) 
from the mode shape and the equivalent generator parameters, and the  $\angle \alpha$
estimated from random load variations with  the method of section \ref{Denominator}.

\begin{figure}
\centering
\subfigure{\includegraphics[width=\columnwidth]{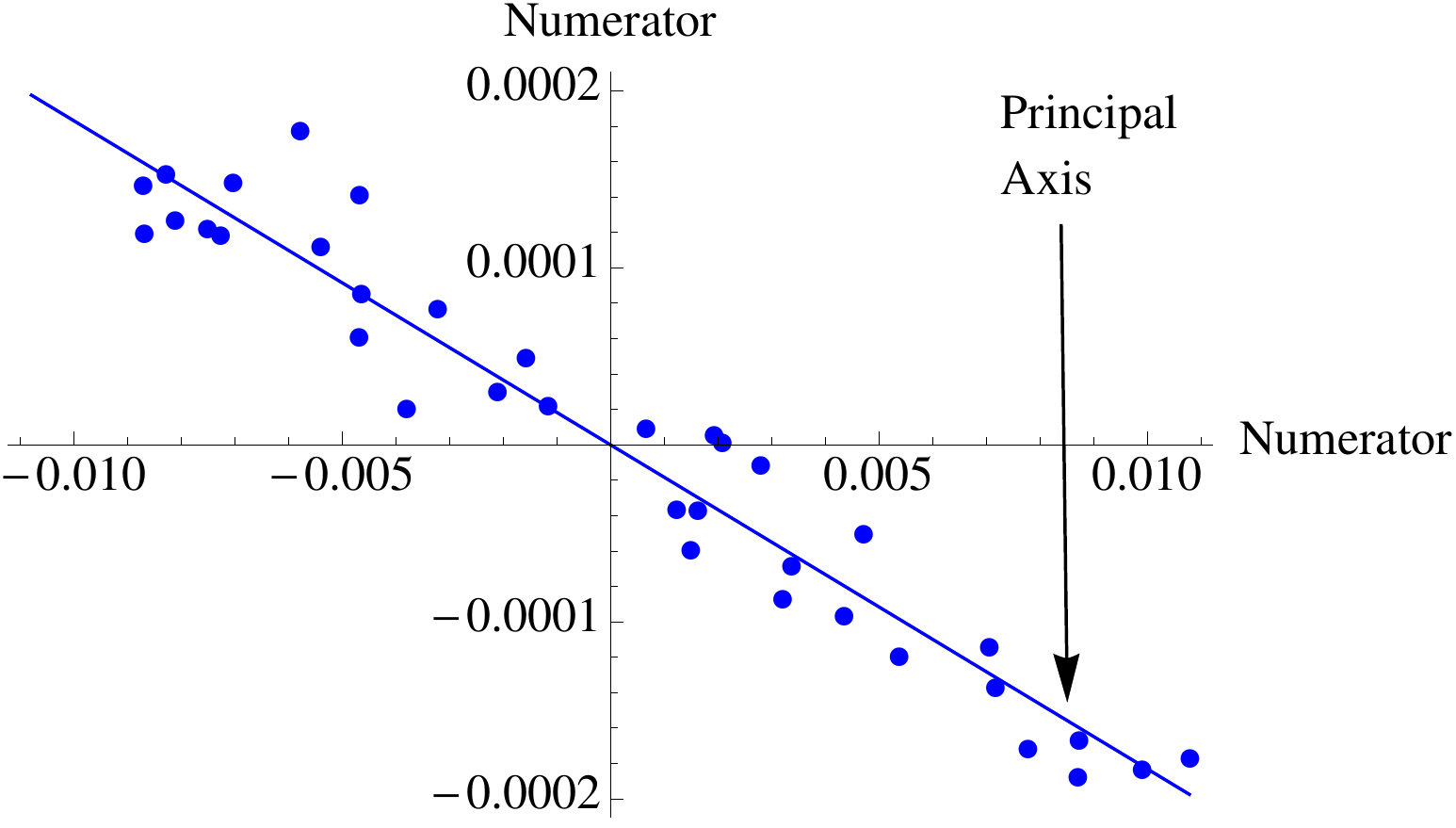}}
\subfigure{\includegraphics[width=\columnwidth]{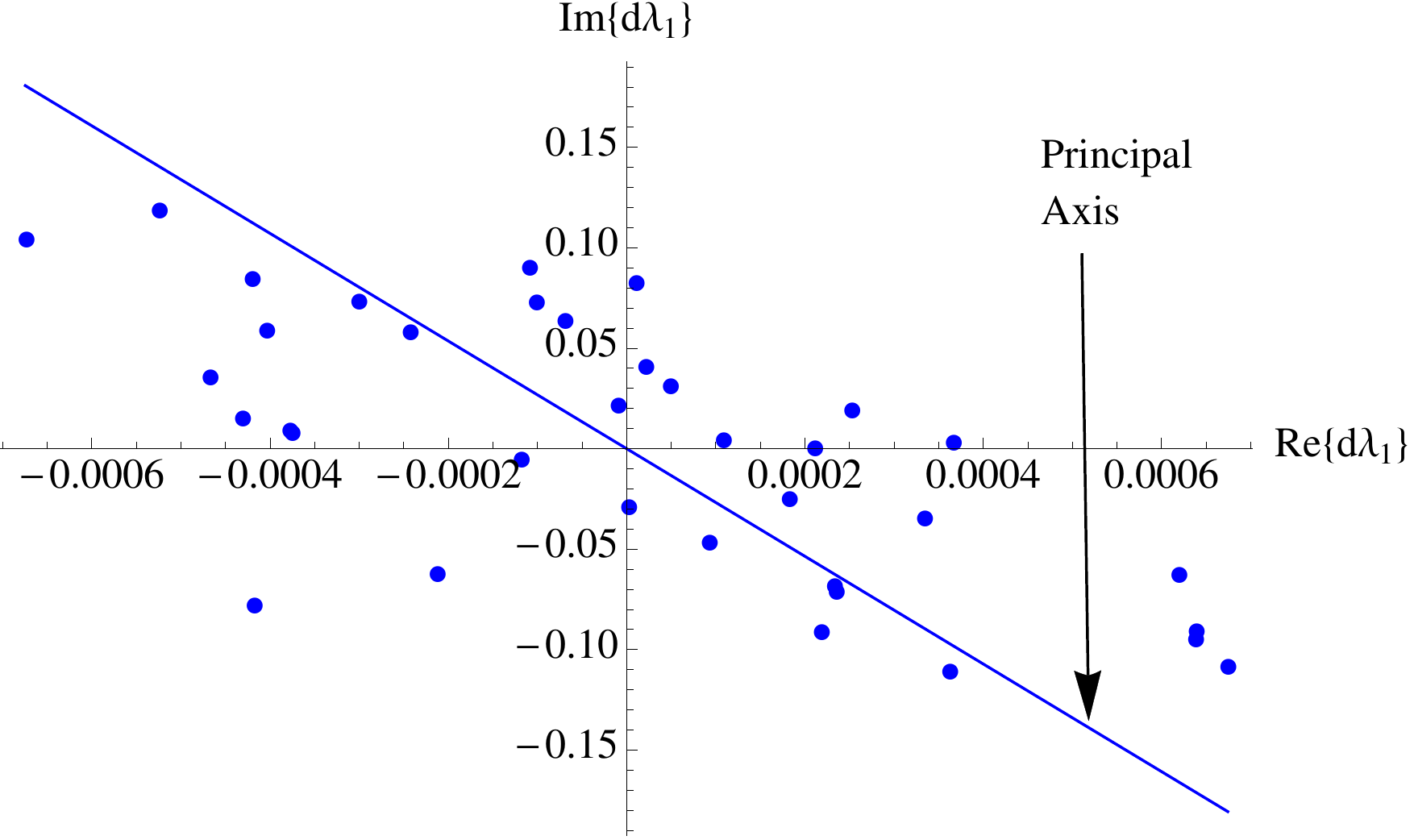}}
\caption{\small \label{Alpha Phase Plots}50  samples  of the numerator of 
(\ref{dlambda}) and $d\lambda$ after trimming by 30\%. Principal axes are
computed and shown as lines. }
\end{figure}

For each interarea mode, the set of random loads used in the computation were generated with the software 
Mathematica. The active power random load vector $P^r$ was sampled from a normal distribution of zero mean\footnote{Load 12 has exceptionally high reactive power and was not varied.}
The components of the reactive power random load vector $Q^r$ were computed as
\begin{align}
\label{Reactive Random Load}
Q^r_i=\frac{P_i}{Q_i}P^r_i.
\end{align} 
$P_i$ and $Q_i$ are the active and reactive power demanded by load  $i$ at the base case.
Then the load flow solution for the vector of loads $P+P^r$ and $Q+Q^r$ were computed and $d\lambda$ and 
the numerator of (\ref{dlambda})  were computed. 50 random load scenarios were generated.
Fig.~\ref{Alpha Phase Plots} shows samples for $\lambda_1$ after being
trimmed by 30\% to remove outliers and analyzed with principal component analysis. 
This 2-dimensional  $\lambda_1$  data was trimmed with multidimensional trimming based on projection 
depth \cite{ZuoStatics}.
Projection depth induces order for high dimensional  data, 
which makes trimming straightforward. The results in Table \ref{AlphaPhase} 
show that the method gives a very good estimation of $\angle \alpha$.
\begin{table}[h]%
\caption{\label{AlphaPhase}{Estimated and exact phases of $\alpha$} for the interarea modes}
\centering
\scalebox{0.95}{
 \begin{tabular}{cccc}
Mode & Exact $\angle \alpha$ & Estimated $\angle \alpha$ & Exact $\angle \alpha$ - Estimated $\angle \alpha$ \\ 
\hline
1 & $88.658^{\circ}$ & 88.718$^{\circ}$ &   $-0.0600^{\circ}$\\
2 & $ 90.257^{\circ}$ & 90.281$^{\circ}$  & $-0.0241^{\circ}$\\
3 & $89.700^{\circ}$ & 89.685$^{\circ}$ & $\phantom{-}0.0156^{\circ}$\\
4 &  $89.026^{\circ}$ & 89.126$^{\circ}$ & $-0.1004^{\circ}$ \\ 
\hline
\end{tabular}
}
\end{table}
\subsection{Ranking generator pairs to damp mode 1}
\begin{figure}[h]
\centering
\includegraphics[width=0.8\columnwidth]{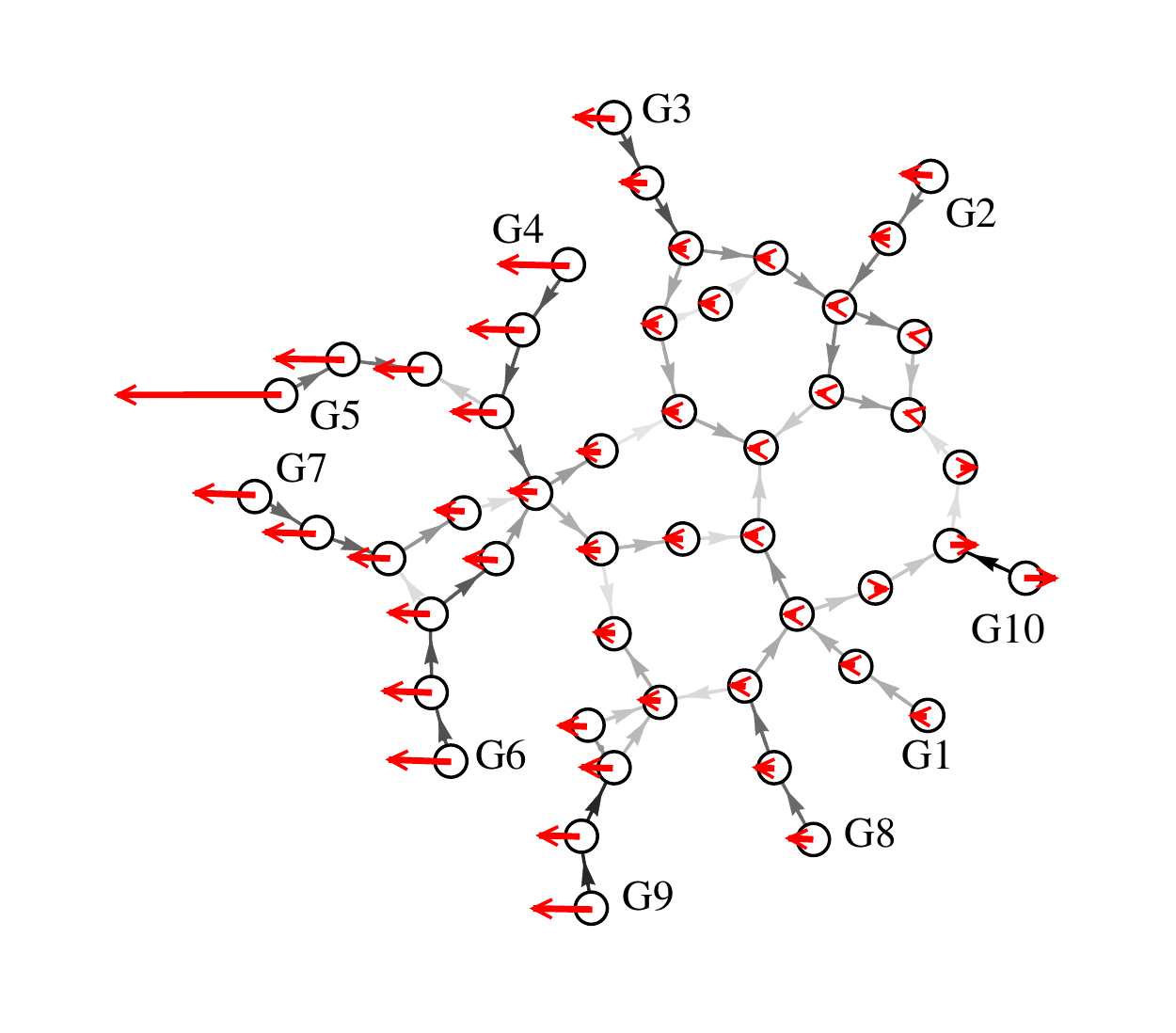}
\caption{\small \label{ModeShape1GraphActivePowerFlowThroughLines}The gray lines joining the buses show the magnitude of the
active power flow with the grayscale and the direction of the active power flow with the arrows. The red arrows at each bus
show the pattern of the oscillation for mode 1; that is, the magnitude and direction of the entries of the right
eigenvector $x_{\delta}$.
}
\end{figure}

Fig.~\ref{ModeShape1GraphActivePowerFlowThroughLines} shows the power flow $p$ and oscillating mode pattern of 
interarea mode 1  at the base case. 
The mode pattern shows that generators G2 through G9 are oscillating against G10. The
component of G10 is not very large compared with the components of the other generators, but G10 
is a large generator that represents an equivalent of the New York State grid. 
From Fig.~\ref{ModeShape1GraphActivePowerFlowThroughLines} we can see that generator G5 participates most in the 
oscillation, G8 has  a small participation in the oscillation, and G1 does not participate in the oscillation.
The method of section \ref{Ranking generator pairs} was applied to rank the generator pairs that 
best increase the damping ratio for a small redispatch of 0.01 pu, and the top 10 generator pairs are shown in 
Table \ref{dZeta1LinearRank}.
The top 9 generator pairs all include G5-; that is, decreasing generation at G5 with increasing generation elsewhere.
The change in damping ratio for these pairs are of the same order of magnitude, so it is clear from Table
\ref{dZeta1LinearRank} that the largest changes in damping ratio are due to the generator pairs involving G5-.
\begin{table}[h]%
\caption{\label{dZeta1LinearRank}Generator pairs ranked by change in $\zeta_1(\%)$; redispatch = 0.01 }
\centering \begin{tabular}{cc|lc}
Generator Pair& $d\zeta_1(\%)$  & Generator Pair & $d\zeta_1(\%)$ \\ 
\hline
1  G6+,G5- &  0.00554  & \hspace{1.5mm}6 G1+,G5- & 0.00505 \\
2  G3+,G5- &  0.00549 &\hspace{1.5mm}7 G9+,G5-& 0.00502 \\
3  G7+,G5- &  0.00545  & \hspace{1.5mm}8 G4+,G5-& 0.00489 \\
4  G2+,G5- & 0.00537    & \hspace{1.5mm}9  G10+,G5-& 0.00481 \\
5  G8+,G5- & 0.00510    & 10 G6+,G10- & 0.000725\\
\hline
\end{tabular}
\end{table}

\subsection{Ranking generator pairs to damp mode  2}
\begin{figure}[h]
\centering
\includegraphics[width=0.8\columnwidth]{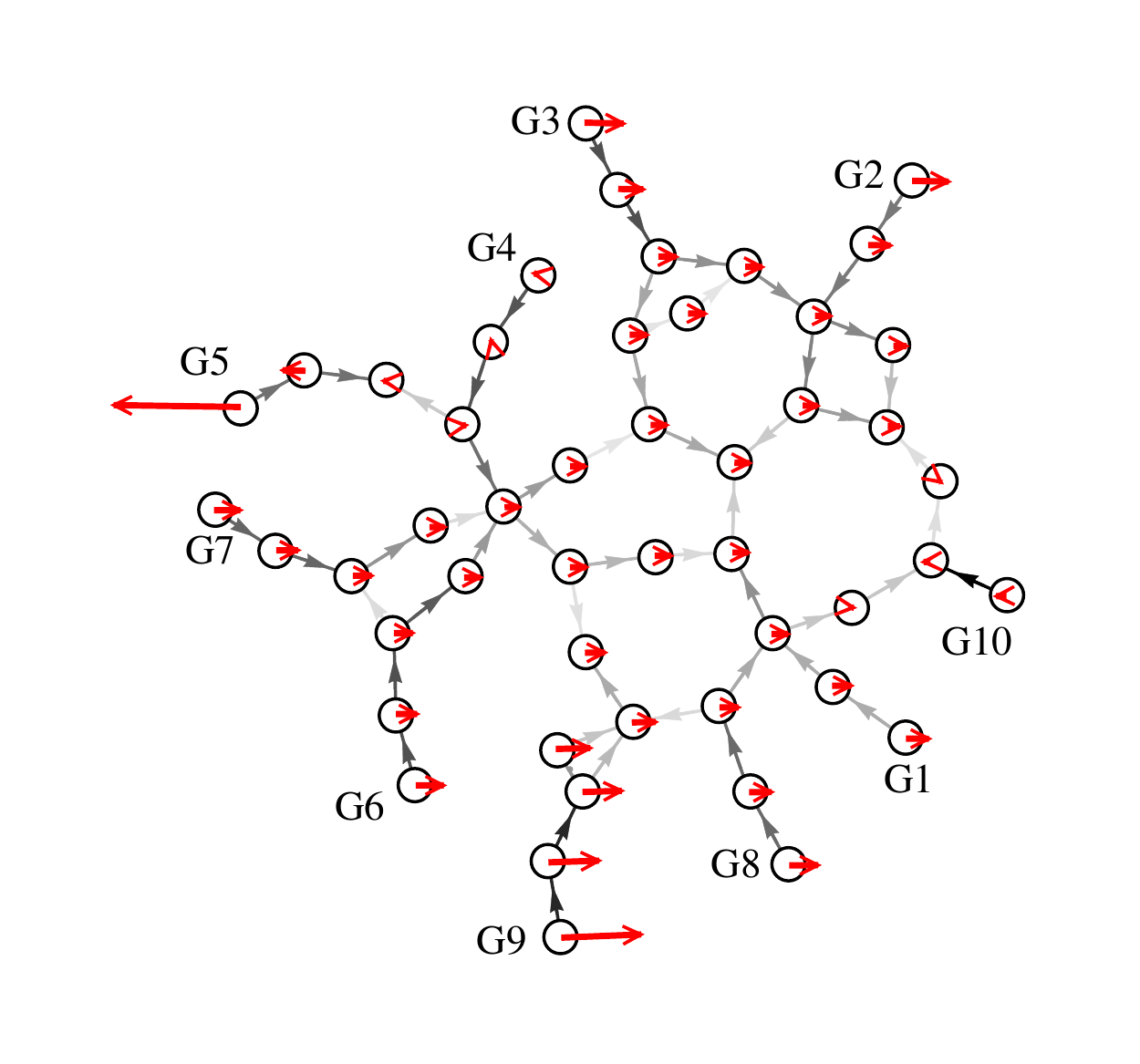}
\caption{\small \label{ModeShape2GraphActivePowerFlowThroughLines}The gray lines joining the buses show the magnitude of the
active power flow with the grayscale and the direction of the active power flow with the arrows. The red arrows at each bus
show the pattern of the oscillation for mode 2.
}
\end{figure}

Fig.~\ref{ModeShape2GraphActivePowerFlowThroughLines} shows the power flow $p$ and oscillating mode pattern of
interarea mode $\lambda_2$ at the base case.
The mode pattern shows that generators G1-G3 and G6-G9 are oscillating against G5.
Generators G5 and G9  participate most in the oscillation. 

The method of section \ref{Ranking generator pairs} was applied to rank the generator pairs 
that best increase the damping ratio for a small redispatch of 0.01 pu, and the top 10 generator pairs are shown in 
Table \ref{DampingRatioRankAnddLambdaMode2}. 
The top 9 generator pairs all include G5+; that is, increasing generation at G5 with decreasing generation elsewhere.
 Although the change in damping ratio is of the same order of magnitude for the 10 pairs, 
the increase in damping ratio for pair 10 G4+,G9-, is roughly half of the increase of damping ratio for pair 9 G5+,G4-.

\begin{figure}
\centering
\subfigure[$|\mbox{Re}\{C_{\theta}\}|$ for $\lambda_2$.]{\includegraphics[width=4.3cm]{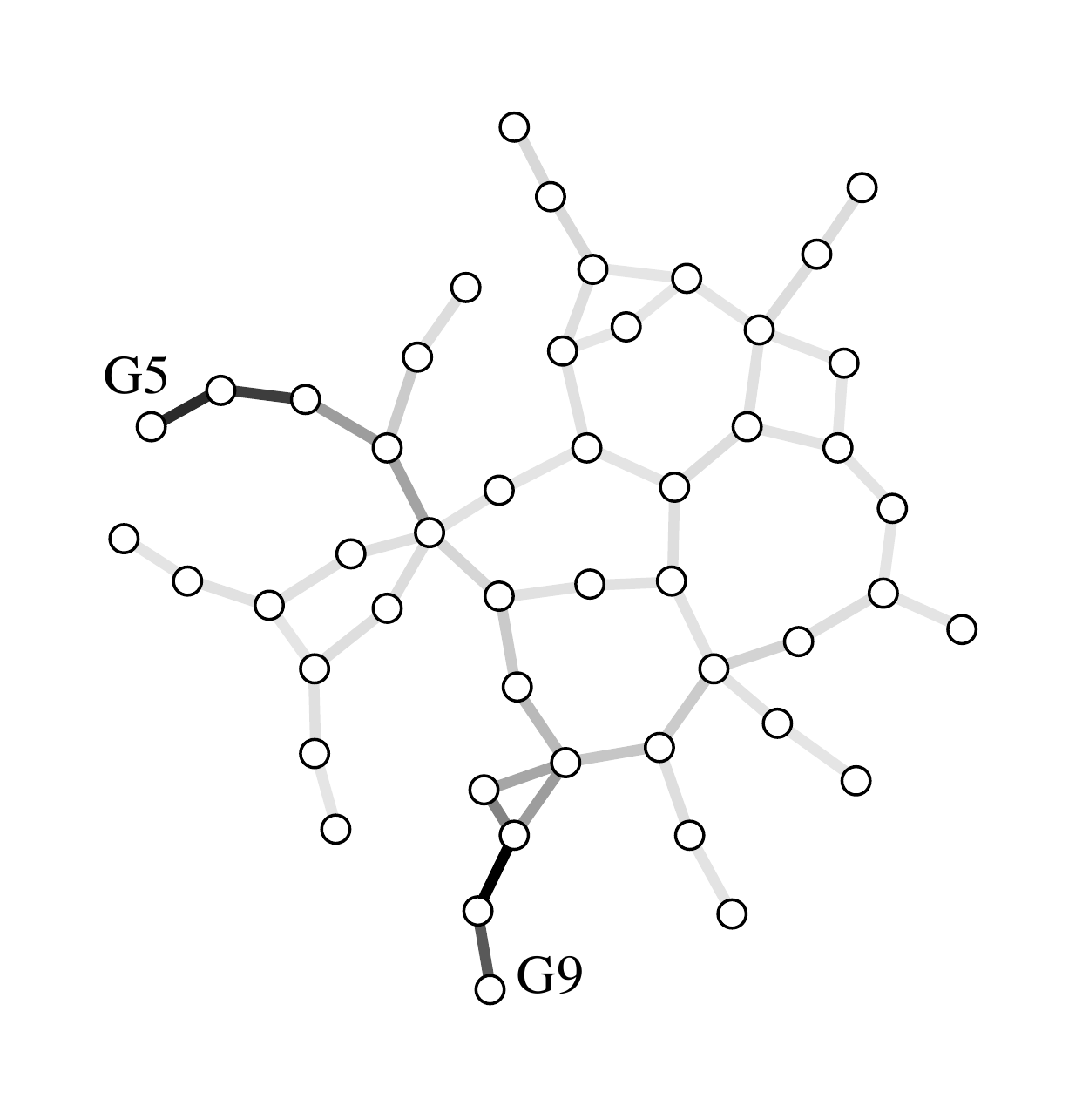}}
\subfigure[$|dp|$ for redispatch G5+,G9-.]{\includegraphics[width=4.3cm]{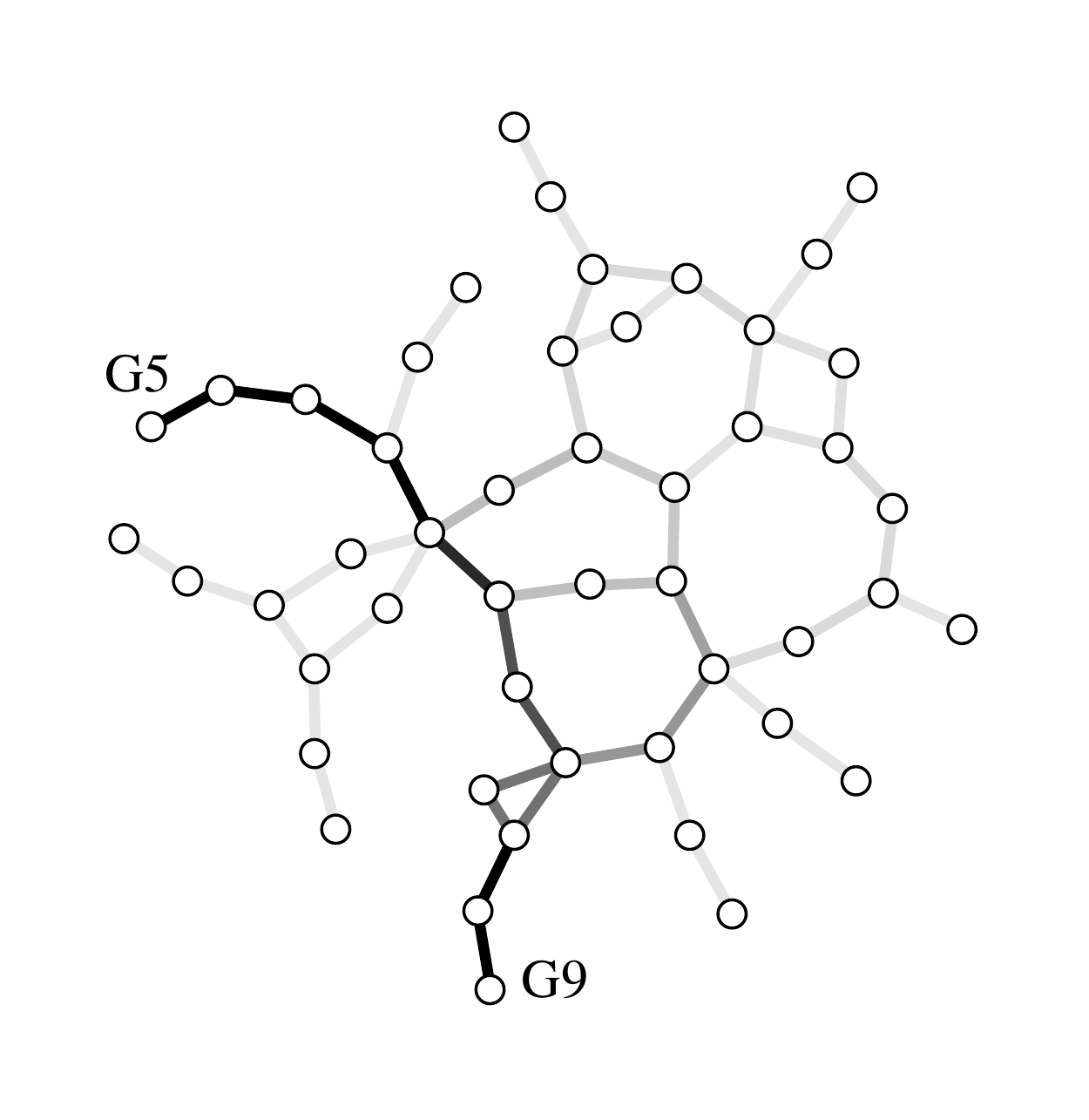}}
\subfigure[$|d\theta|$ for G5+,G9-.]{\includegraphics[width=4.3cm]{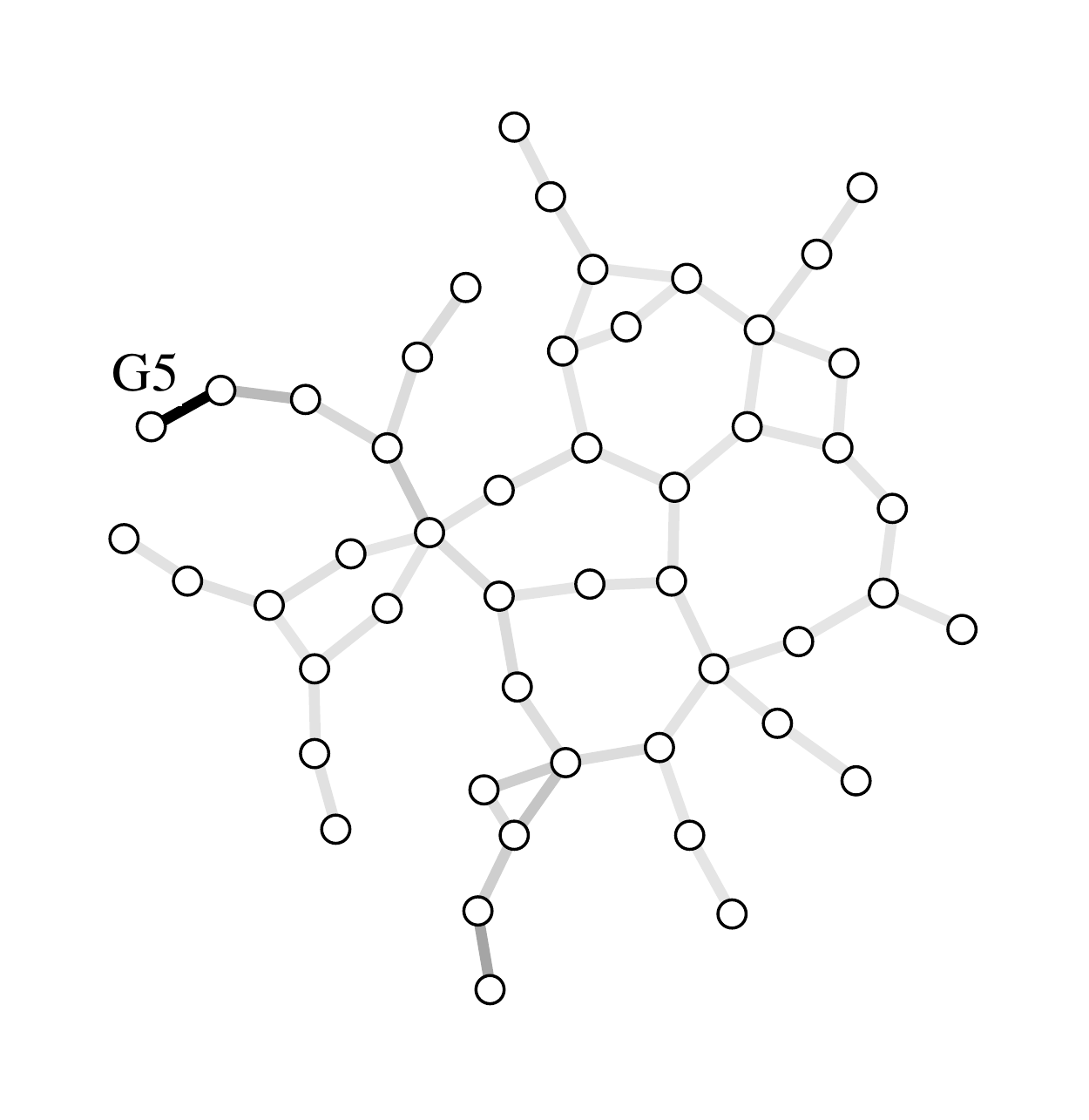}}
\subfigure[$|\mbox{Re}\{C_{\theta}\}\cdot d\theta|$ for  $\lambda_2$ \& G5+,G9-.]{\includegraphics[width=4.3cm]{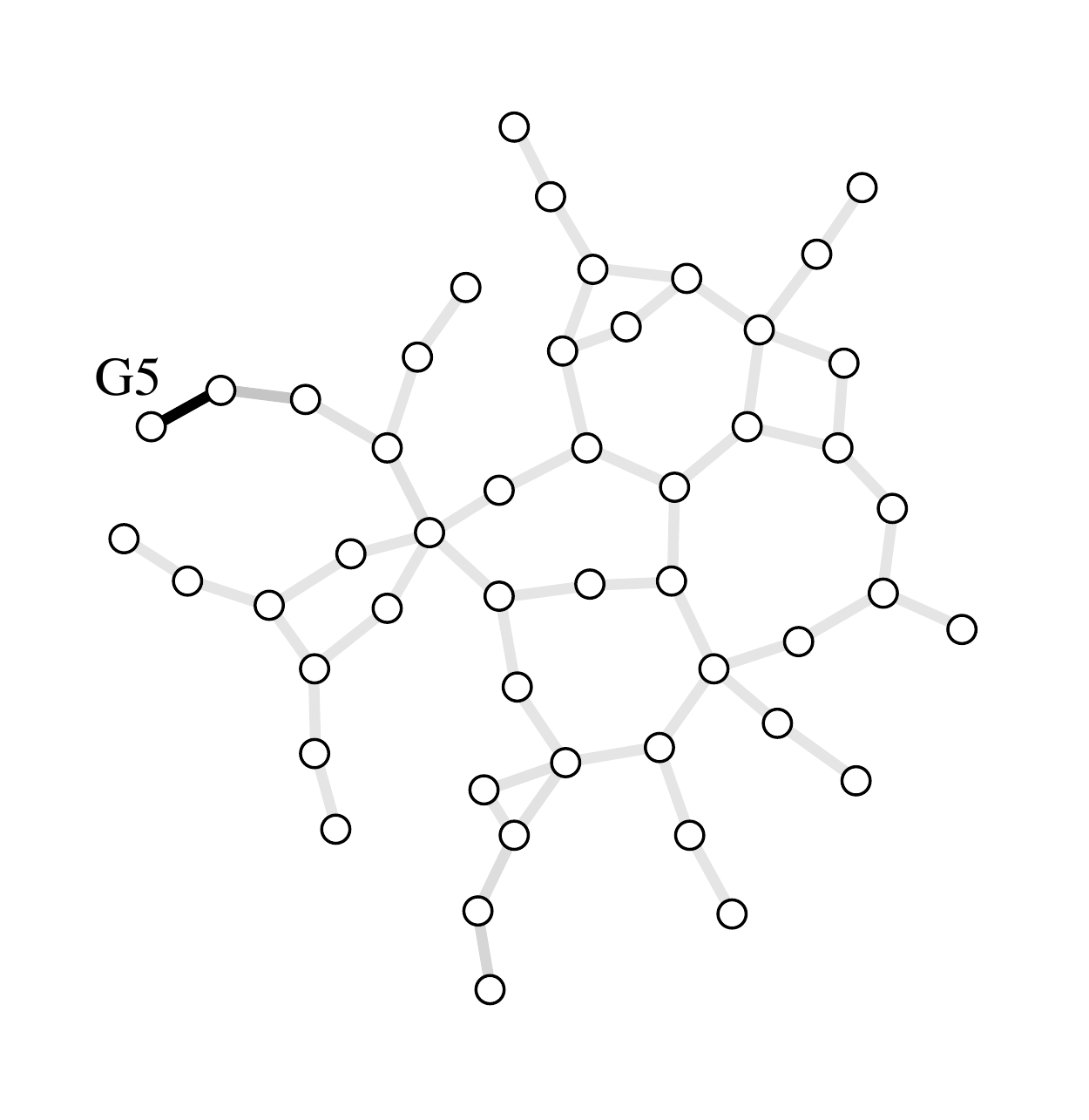}}
\caption{\label{G5G9Figs} Gray scale in the lines shows the components
of the specified vector.}
\end{figure}

Now we analyze the most significant components of  formula  (\ref{dlambda})  to explain why G5 is playing a key role in damping mode~2.
\begin{table}[h]
\caption{\label{DampingRatioRankAnddLambdaMode2}Generator pairs ranked by change in $\zeta_2(\%)$; redispatch = 0.01 }
\centering \begin{tabular}{lccccc} 
&&&& \hspace{-2mm}Re$\{C_{V}\}$ \\ 
\hspace{1mm}Gen. Pair&$d\zeta_2(\%)$& $d\lambda$ &\hspace{-2mm}Re$ \{C_{\theta}\}\cdot d\theta$ 
& \hspace{-2mm}$\cdot \,dV$ \\ 
\hline
1  G5+,G9- &9.60E-3 &-4.23E-4 - j0.008 & -3.04E-4  & -1.19E-4  \\
2 G5+,G8-& 7.30E-3 & -3.06E-4 - j0.010 & -2.23E-4& -8.32E-5 \\
3 G5+,G1-& 7.28E-3&-3.02E-4 - j0.011 & -2.21E-4 & -8.12E-5 \\
4 G5+,G10-&7.26E-3&-3.00E-4 - j0.011 & -2.19E-4 & -8.10E-5\\
5 G5+,G2-& 6.85E-3&-2.85E-4 - j0.010  & -2.10E-4   & -7.55E-5\\ 
6 G5+,G7-&6.79E-3&-2.84E-4 - j0.010  & -2.11E-4  &-7.30E-5 \\
7 G5+,G6-&6.77E-3&-2.84E-4 - j0.010 &-2.11E-4    &-7.28E-5\\
8 G5+,G3-&6.74E-3&-2.81E-4 - j0.010  &-2.08E-4   &-7.38E-5 \\
9 G5+,G4-&6.05E-3&-2.52E-4 - j0.009  &-1.90E-4   &-6.21E-5\\
\hspace{-1.5mm}10 G4+,G9-&3.53E-3&-1.71E-4 - j0.001  &-1.14E-4   &-5.66E-5 \\
\hline
\end{tabular}
\end{table}
Table \ref{DampingRatioRankAnddLambdaMode2}  shows that the pairs with the largest change in damping ratio $d\zeta$ are  the
ones with the largest increase in the damping Re$\{d\lambda\}$, so we can focus on the real part of (\ref{dlambaCcoefficients}). Moreover,
Table \ref{DampingRatioRankAnddLambdaMode2} also shows that the changes Re$ \{C_{\theta}\}\cdot d\theta$ are larger than 
the changes Re$\{C_{V}\}\cdot dV$, so we focus on analyzing the terms of (\ref{dlambaCcoefficients}) related to $d\theta$:
\begin{align}
\hspace{-2mm}\mbox{Re}\{d\lambda\} 
\label{RedLambda}
 =\mbox{Re}\{C_{\theta}\}\cdot d\theta +\mbox{Re}\{C_{V}\}\cdot dV\approx \mbox{Re}\{C_{\theta}\}\cdot d\theta
\end{align}
Fig.~\ref{G5G9Figs} shows different quantities related to the 56 lines of the New England system in gray scale.
Each generator is represented in the network by an internal bus and a terminal bus.
There are two lines in series at the edge of the network associated with each generator. 
The line joining the internal bus to the terminal bus represents the generator transient reactance, and the other line 
lumps together the transformer and lines joining the generator terminal bus to the network.
(G10 differs since it represents New York state.)
Fig.~\ref{G5G9Figs}(a) shows  $|\mbox{Re}\{C_{\theta}\}|$, the absolute value of $d\theta$'s coefficient in (\ref{RedLambda}). 
The lines that represents the transient reactance of G5 and the transient reactance of G9 have large components of $|\mbox{Re}\{C_{\theta}\}|$.
Fig.~\ref{G5G9Figs}(b) shows $|dp|$, the absolute value of the change in power flow in lines  for the
best ranked generator pair G5+,G9-. 
 As expected, several lines have a significant change in power flow, but Fig.~\ref{G5G9Figs}(c) shows that 
only the line that represents the transient reactance of G5 has a large change in angle $|d\theta|$. This is due to the fact that transient reactance of  G5
is much larger than the 
reactance of any of the other 55 lines of the system (see appendix A). 
So, for any possible generator pair that involves G5, the line
that represents the transient reactance of G5 will always have the largest change in angle. This large change in angle, combined with a large
coefficient $|\mbox{Re}\{C_{\theta}\}|$, produces the dominant term of (\ref{RedLambda}), as shown by 
Fig.~\ref{G5G9Figs}(d). Thus  G5 is the key generator to participate in redispatch for producing the largest changes in
damping  for mode 2.
Although Fig.~\ref{G5G9Figs}(a) shows that there are other lines that have large $d\theta$'s real coefficient, and Fig.~\ref{G5G9Figs}(b)
shows that other lines have an important change in power $|dp|$, Fig.~\ref{G5G9Figs}(c) shows that such lines do not have a large change in angle
$d\theta$, and as a result their associated terms in (\ref{RedLambda}) are
not large, as seen in Fig.~\ref{G5G9Figs}(d). 
This analysis shows a mechanism of how damping by redispatch works by changing the angles across lines that have large coefficients $C_{\theta}$.
\subsection{Larger redispatches}
\begin{figure}
\centering
\subfigure{\includegraphics[width=9cm]{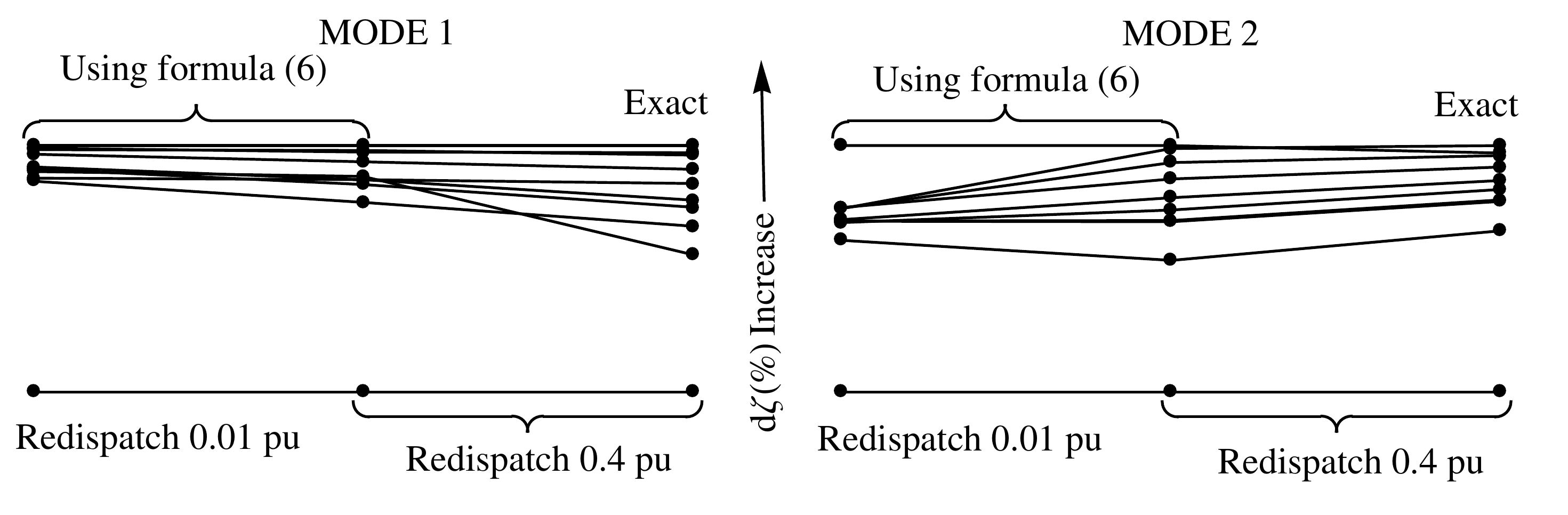}}
\caption{\label{FigureFor3Ranks} Changes in damping ratio for Mode 1 and Mode 2 for the top 10 generator pairs from Tables \ref{dZeta1LinearRank}, 
\ref{DampingRatioRankAnddLambdaMode2}, \ref{dZetaRankLargerRedispatch}  rescaled to the same range to allow comparison. 
For each mode, left hand dots use formula (\ref{dlambda}) for a small redispatch, middle dots  use  (\ref{dlambda}) for a larger redispatch,
and the right hand dots are the exact calculation for the larger  redispatch. Clustering of similarly effective generator pairs is shown by close dots and changes in ranking 
appear as lines crossing.
}
\label{ranking}
\end{figure}

 Larger redispatches introduce some nonlinearity in the change in the damping ratio. This nonlinearity 
 arises from three sources: the change in the load flow, the change in the eigenvalue $d\lambda$, and the change in the damping ratio.
 For a given size of redispatch of a generator pair, we can compute the exact nonlinear change in the load flow due to the redispatch, use (\ref{dlambda}) to linearly estimate the change in the eigenvalue based on measurements, and then 
 compute the nonlinear change in the damping ratio. This is a large signal  application of (\ref{dlambda}),
 and the  top 10 generator pairs for modes 1 and 2 are shown in Table \ref{dZetaRankLargerRedispatch} for a  redispatch of 0.4 pu.
 For comparison, Table \ref{dZetaRankLargerRedispatch}  also shows the exact calculation that uses the nonlinear computation of $d\lambda$.
 As shown in Fig.~\ref{ranking}, the use of  (\ref{dlambda}) for a larger redispatch gives  the same grouping of the top 9 effective generator pairs and a 
 similar  ranking and grouping of the top 10 generator pairs as the  exact calculation.
 An exception is that for mode 1, the 5th ranked generator pair G9+,G5- for a larger redispatch using  (\ref{dlambda}) becomes the 9th ranked generator pair for the exact calculation.

 Fig.~\ref{ranking} also compares the changes in the damping ratio of the top 10 generator pairs for a small redispatch using (6) to the larger redispatch using (6).
 The grouping of the top 9 effective generator pairs is preserved and the approximate ranking is preserved.

Overall, some details of the rankings differ for similarly effective generator pairs, but since 
the ranking will be used to provide a set of effective generators pairs from which 
an operationally suitable pair can be selected for redispatch by operators, 
the performance of the ranking is satisfactory.
\begin{table}[h]%
\caption{\label{dZetaRankLargerRedispatch}Generator pairs ranked by change in $\zeta(\%)$; redispatch = 0.4 }
\centering \begin{tabular}{lcc|lccc}
~MODE 1  &  \multicolumn{2} {c|} {~~$d\zeta_1(\%)$} & ~MODE 2 & \multicolumn{2} {c} {~~$d\zeta_2(\%)$}\\
\hspace{1mm}Gen. Pair & Using (\ref{dlambda})& Exact&\hspace{1mm}Gen. Pair &Using (\ref{dlambda}) & Exact \\ 
\hline
1   G6+,G5- &0.181 & 0.138  &1 G5+,G9- &0.563  &0.600 \\
2   G7+,G5- &0.177 & 0.134 &2 G5+,G10- &0.558 &0.615 \\
3   G3+,G5- &0.177 &  0.135 &3 G5+,G1- &0.534 &0.595 \\
4   G2+,G5- &0.171 &  0.128 &4 G5+,G8- &0.506 &0.572  \\
5  G9+,G5- &0.162 &  0.092 &5 G5+,G2-  & 0.473&0.545  \\ 
6   G4+,G5- & 0.160& 0.122 &6 G5+,G3- &0.453 &0.527 \\
7 G8+,G5- &0.159 &  0.115  &7 G5+,G7- &0.434 &0.506\\
8 G1+,G5- &0.155& 0.112  &8 G5+,G6- &0.432  &0.503\\
9 G10+,G5- &0.145&  0.104  &9 G5+,G4- &0.365 &0.444 \\
\hspace{-1.5mm}10 G6+,G10- &0.030 & 0.034 &\hspace{-1.5mm}10 G4+,G9- &0.139 &0.120\\
\hline
\end{tabular}

\end{table}

\section{Discussion of generator modeling}
\label{discussgenmodels}

Since our approach depends in new ways on both measurements and  an equivalent second-order swing equation generator model,
our generator modeling requirements are different than in other approaches to suppressing oscillations.
Section~\ref{genmodeling} explains that the generator dynamics are approximated by a swing equation, but we do not need to determine the parameters of 
the swing equation for any individual generator. 
This section 
further discusses the generator modeling. 

As a general observation, in closed loop control of oscillations with power system stabilizers, which forms most of the literature on suppressing oscillations, 
the generator and its controls  need to be modeled in sufficient detail. Indeed the designed control gains directly affect entries of the Jacobian to damp the oscillatory mode. 
Our control is open loop and works by the entirely different principle of exploiting system nonlinearity by changing the operating point at which the Jacobian is evaluated. 
This changes the focus from the linear parts of the model to the nonlinearities. 

Formula (\ref{dlambda}) computes the first order sensitivity of the oscillatory mode eigenvalue to generator redispatch. Appendix~B proves that the first order eigenvalue sensitivity to redispatch does not depend on linear parts of the power system model. In particular, 
if the generator magnetic saturation and hysteresis are neglected, the higher order parts of the generator modeling are linear and the eigenvalue sensitivity only depends on the nonlinearity in the swing equation and any stator algebraic equations and does not depend on the linear higher order part of the generator dynamic modeling. 
This does suggest that  the linear higher order generator dynamics can be omitted  in deriving the formula.

In applying formula (\ref{dlambda}), we do not use a model of the power system dynamics. Instead we rely on measurements of the system dynamics, particularly the eigenvalue and the right eigenvector of the mode and the phase of the complex scalar parameter $\alpha$ that combines together all of the generator dynamics. There are no model assumptions in these measured quantities. That is, if part of the power system affects the oscillatory dynamics, the effect will appear via the measurements used by the formula.

We did an initial test of the approximation involved in generator modeling with the interarea mode of the 3-generator model similar to  Fig.~1 with a sixth-order generator model at each bus.
Formula  (\ref{dlambda})  is applied to this detailed model by measuring the phase of $\alpha$, and using the computed mode, mode shape, and load flow to estimate the change $d\lambda$
in the eigenvalue that would arise from small redispatches in generation.
Then the detailed model is used to compute the exact change  $d\lambda$ in the eigenvalue.
The comparison of the approximate and exact $d\lambda$  is shown in Table \ref{SixOrderModelVSSwingModel}.
Similarly to the intended application of the formula to ranking redispatches in a real power system, the power system model with sixth-order generators does not have 
the parameters of the equivalent second order generator models available. That is, the magnitude of $\alpha$ is not known, and only the phase of $\alpha$ is estimated,
and so the formula predicts $d\lambda$ to within a constant real multiplier.
Therefore in Table \ref{SixOrderModelVSSwingModel}  we compare the ratios of $|d\lambda|$ for each redispatch to $|d\lambda|$ for the redispatch of generators G1 and G2, as well as 
comparing the phases of $d\lambda$.
The approximation of $d\lambda$ in Table \ref{SixOrderModelVSSwingModel} is close enough to be acceptable for ranking of generator redispatches.

While the generator modeling issues should be investigated further in future work, and further analytic progress is not ruled out, 
both theoretical considerations of the irrelevance of linear parts of the generator model and an initial test 
indicate that combining measurements of the dynamic quantities with a formula assuming a second order swing equation can be adequate for 
ranking generator redispatches.

\begin{table}[h!]
\caption{\label{SixOrderModelVSSwingModel}{Eigenvalue changes for redispatch of 0.01 pu in 3-generator system with sixth-order generator modeling} }
~\\
\centering \begin{tabular}{c|cc|cc}
Generator &  \multicolumn{2} {c|} {Exact}  &\multicolumn{2} {c} {Approximate with formula}  \\
pair& $\angle d\lambda$ & $|d\lambda|$ ratio& $\angle d\lambda$ & $|d\lambda|$ ratio \\
\hline
G1+,G2-&$-109.5^{\circ}$ & 1.00 & $-106.3^{\circ}$& 1.00 \\
G2+,G3-&~$-98.6^{\circ}$ & 2.54 & ~$-98.5^{\circ}$ & 2.12\\
G1+,G3- &$-101.7^{\circ}$ & 3.53 &$-101.0^{\circ}$& 3.11\\
\end{tabular}
\end{table}

\section{Conclusions}
There has been success in monitoring interarea modal damping with synchrophasor measurements \cite{PierrePS97,WiesPS03};
the next step is to leverage synchrophasor measurements to provide advice to the operators to
maintain a suitable modal damping ratio when the damping ratio is insufficient.
Difficulties in accomplishing  this in the past include the lack of wide-area online dynamic models and 
standard formulas that depend on quantities that cannot be measured.
In this paper, we circumvent these difficulties  by 
calculating the best generator pairs to redispatch to maintain 
modal damping by combining synchrophasor  and state
estimator measurements with a new analytic formula
for the sensitivity of the mode eigenvalue with respect to
generator redispatch.
The assumed equivalent generator dynamics only appears as a complex factor $1/\alpha$ common to all redispatches
and we propose a method of estimating the phase of $\alpha$ from ambient measurements.
The new formula is somewhat complicated, and we explain and illustrate how it works in 3 and 10 generator examples. 
Future work may well discover further insights and applications using the formula.
In summary, we make substantial progress towards practical application of a new formula to damp 
interarea oscillations based on measurable quantities.

\footnotesize
\renewcommand{\em}{\null}


\section*{Appendix A: New England generator data}
\label{New England Machine Data}
\begin{table}[h!]
\centering
 \begin{tabular}{ccccccc}
Gen. & Terminal & Internal& & & & $V$ Internal\\
No. & Bus No. & Bus No.&$h$ [s] & $d$ [s]  & $x'_{d}$ & Bus \\
\hline
1 &30 &40& 42.0 &    0.0267     & 0.031 & 1.0501 \\
 2 & 31 &41&30.3 &    0.0161    & 0.0697 & 1.0388 \\
 3 & 32 & 42& 35.8 &  0.0209    & 0.0531 & 1.0439 \\
 4 & 33 &43& 28.6 &   0.0243    & 0.0436 & 1.0348 \\
 5 & 34 & 44&26.0 &   0.0014      & 0.132 & 1.2098 \\
 6 & 35 & 45&34.8 &   0.0277    & 0.05 & 1.0941 \\
 7 & 36 & 46&26.4 &   0.0140    & 0.049 & 1.0944 \\
 8 & 37 & 47&24.3 &   0.0116      & 0.057 & 1.0705 \\
 9 & 38 & 48&34.5 &     0.0002     & 0.057 & 1.1252 \\
 10 & 39 & 49&500.0 & 0.3979    & 0.006 & 1.0317\\ 
\hline
\end{tabular}
\end{table}
\vspace{-10pt}
\noindent
$V_i$ is the internal constant voltage
magnitude of generator $i$ at the base case; i.e., at zero redispatch, and $x'_{d}$ is the transient generator reactance.

\normalsize

\section*{Appendix B: Irrelevance of linear modeling}
\label{irrelevance}
This appendix proves from (\ref{ReportFormula})  that the first order eigenvalue sensitivity to redispatch does not depend on linear parts of the power system model.
This result was first mentioned in \cite[section 4.6]{DobsonPSERC99}.
It is convenient to use
the extended differential-algebraic form of the power system equations \cite{SmedPS93} with state vector $z$,
Jacobian $\bar J$, and  extended right and left eigenvectors $\bar v$ and $\bar w$.  
In this notation, (1) becomes 
\begin{align}
\label{ReportFormulabar}
\frac{\partial \lambda}{\partial p} = \frac{\bar w\bar J_p\bar v}{\bar w\bar  v}
\end{align}
In this case, the parameter $p$ parameterizes the generator redispatch, and it appears linearly in the 
system equations. Therefore $p$ does not appear explicitly in the Jacobian $\bar J$, and 
\begin{align}
\bar J_p=\frac{\partial \bar J}{\partial p} =
\sum_{k}\frac{\partial \bar J}{ \partial z^k}z_{p}^k
\end{align}
where $z_{p}=\frac{\partial z}{\partial p}$ is the sensitivity of the operating point $z$ to the redispatch.
Then substituting in (\ref{ReportFormulabar}) and writing it out in coordinates gives
\begin{equation}
\frac{\partial \lambda}{\partial p} =
\frac{
\displaystyle\sum_{i,j,k}\bar w^i
\frac{\partial \bar J^{ij}}{\partial z^k}
z_{p}^k\bar v^j
}{
\displaystyle \sum_{i}\bar w^i\bar v^i}
\label{lambdap}
\end{equation}
It is clear from (\ref{lambdap})  that the linear parts of the model vanish in
$\frac{ \partial \bar J^{ij}}{\partial z^k}$ and that if a dynamic state is associated with a linear differential equation,
the corresponding entry of $\bar w$ gets multiplied by zero in the numerator 
of (\ref{lambdap}).

\vspace{4mm}

\footnotesize
\noindent
{\bf Sarai Mendoza-Armenta} 
(M 13) received the PhD in Physics from Instituto
de F\'{\i}sica y Matem\'aticas, Universidad Michoacana, Mexico in 2013.
She was visiting scholar at Iowa State University from 
March 2012 to March 2013.
She was post-doctoral research associate in
the Electrical and Computer Engineering Department 
at Iowa State University.
 \\

\noindent
{\bf Ian Dobson}
(F 06) received  the BA  in Maths
from Cambridge University  and the  PhD  in Electrical Engineering
from Cornell University. 
He previously worked for British industry and  the University of Wisconsin-Madison and is currently Sandbulte professor of  engineering at Iowa State University.

\end{document}